\documentclass [11pt]{article}
\usepackage[dvips]{graphicx}
\usepackage{subcaption}

\usepackage{epsfig}
\usepackage{graphicx}
\usepackage{inde ntfirst}
\usepackage{amsmath}
\usepackage{amsfonts}
\usepackage{amssymb}
\usepackage{hyperref}
\usepackage{latexsym}

\usepackage{titlesec}

\titleformat{\section}
{\centering\normalfont\Large\bfseries}
{\thesection}{1em}{}

\titleformat{\subsection}
{\centering\normalfont\large\bfseries}
{\thesubsection}{1em}{}

\usepackage{color}
\renewcommand{\thesubsection}{\Alph{subsection}}

\renewcommand{\thesection}{\arabic{section}}

\begin{document}

\begin{center}
\begin{flushright}
	\begin{small}    
\end{small}
 \end{flushright} \vspace{1.5cm}
\Large{\bf Gravitational baryogenesis in scalar-nonmetricity $f(Q,\phi)$ gravity  }
\end{center}

\begin{center}
F. Mavoa $^{(a,b)}$\footnote{e-mail: ferdinand.mavoa@univ-labe.edu.gn, maferdson@yahoo.fr},
M.C.Sow $^{(a)}$\footnote{recteur@univ-labe.edu.gn},
M. G. Ganiou $^{(c)}$\footnote{ganiou.gbenga@uganc.edu.gn}, 
C.S.Tour\'e $^{(a)}$\footnote{	cheikhsalioutoure@gmail.com},
$^{(d)}$ C.R.Tefo \footnote{teffcas@gmail.com
	}

\vskip 4mm
$^{a}$\, {\it  D\'epartement Energie Photovolta\"{i}que, Universit\'e de Lab\'e, R\'epublique de Guin\'ee}\\ 

{\it B.P:(+224) 2010,  Lab\'e, R\'epublique de Guin\'ee}\\

$^b$ \,{\it International Chair of Mathematical Physics and Application (ICMPA) , University d'Abomey-Calavi, B\'enin}\\
 {\it  072 BP 50, Cotonou, B\'enin}\\
 
$^{c}$\, {\it  D\'epartementde Phsique , Universit\'e Gamal Abdel Nasser, R\'epublique de Guin\'ee}\\
{\it BP 1147 Conakry , R\'epublique de Guin\'ee}\\

$^d$ \,{\it D\'epartement de Physique, Facult\'e des sciences techniques Universit\'e de N'z\'er\'e\'ekor\'e, }\\
{\it BP 50 N'z\'er\'e\'ekor\'e,  République de Guin\'e}\\

\vskip 2mm
\end{center}


\begin{abstract}
	In this work, we investigate gravitational baryogenesis in the framework of scalar–nonmetricity theories by considering two classes of modified gravity models, namely $f(Q,\phi)=Q+\xi Q \phi^2$ and $f(Q,\phi)=\alpha Q^n + \beta \phi Q$. These models extend the standard $f(Q)$ gravity through the inclusion of nonminimal couplings between the scalar field and the nonmetricity scalar, leading to nontrivial modifications of the cosmological dynamics. We analyze the evolution of the baryon-to-entropy ratio $\frac{n_B}{s}$ in terms of the cosmic expansion parameter $\gamma$, assuming a power-law behavior of the scale factor. For the first model, we show that $\frac{n_B}{s}$ decreases monotonically with increasing $\gamma$, reflecting the impact of the expansion rate on the efficiency of baryogenesis. The observed baryon asymmetry, $\frac{n_B}{s} \sim 10^{-11} - 10^{-10}$, is successfully reproduced for $\gamma \approx 0.2$--$0.3$ without requiring fine-tuning of the model parameters. For the second model, we explore the parameter space of $\alpha$ and $\beta$, and demonstrate that the correct order of magnitude of the baryon asymmetry can be achieved for physically reasonable values of the parameters. In particular, we find that $\frac{n_B}{s}$ lies within observational bounds for $\alpha \sim 10^{-3} - 10^{-2}$ and $\beta \sim 10^{-2} - 10^{-1}$, with specific combinations yielding excellent agreement with observations. Overall, our results show that scalar–nonmetricity gravity provides a viable and robust framework for explaining the origin of the baryon asymmetry of the Universe. The interplay between nonlinear geometric terms and scalar field couplings plays a crucial role in controlling the baryogenesis mechanism, opening new perspectives in the study of modified gravity and early-Universe cosmology.

\hspace{0,2cm} 
        
\end{abstract}


\section{Introduction}
Since the prediction and experimental discovery of antiparticles \cite{Anderson1933}, it has been widely accepted that the fundamental laws of physics exhibit a high degree of symmetry between matter and antimatter. However, this theoretical symmetry is in striking tension with astrophysical and cosmological observations, which indicate that the observable Universe is overwhelmingly dominated by matter, with a negligible amount of primordial antimatter.

This asymmetry is strongly supported by several independent observations, including the successful predictions of Big Bang Nucleosynthesis (BBN) \cite{Burles2001}, the high-precision measurements of the cosmic microwave background (CMB) \cite{Bennett2003,Spergel2003}, and the absence of significant gamma-ray signals from matter–antimatter annihilation \cite{Cohen1998}. Despite these successes, the origin of the baryon asymmetry of the Universe remains one of the fundamental open problems in modern cosmology and particle physics.

Various baryogenesis mechanisms have been proposed to explain the observed excess of matter over antimatter \cite{Riotto1999,Dine2003,Alexander2006,Mohanty2006,Li2004,Lambiase2013,Oikonomou2015,Oikonomou2017,Pizza2015}. These mechanisms may operate during either the radiation-dominated or matter-dominated eras. The existence of processes violating charge conjugation (C) and charge-parity (CP) symmetries implies a fundamental distinction between matter and antimatter. This opens the possibility for dynamical processes that preferentially generate matter, although a direct quantitative explanation based solely on the observed CP violation is still lacking.

In thermal equilibrium, such as in the early hot Universe, the number densities of particles and antiparticles remain nearly identical. Only when the Universe expands and cools, leading to a departure from equilibrium, can small microscopic asymmetries be amplified into a macroscopic baryon asymmetry.

The necessary conditions for generating such an asymmetry—namely baryon number violation, CP violation, and departure from thermal equilibrium—were first formulated by Sakharov \cite{Sakharov1967}.

In an attempt to establish a connection with dark energy, several studies \cite{Li2002,Li2003} have explored models of spontaneous baryo(lepto)genesis involving interactions between dynamical scalar fields associated with dark energy and ordinary matter. More recently, Davoudiasl et al. \cite{Davoudiasl2004} proposed a mechanism for generating baryon asymmetry even in thermal equilibrium through a dynamical violation of CP symmetry during the expansion of the Universe.

In this framework, CP violation arises from an interaction term coupling the derivative of the Ricci scalar $R$ to the baryon current $J^\mu$, given by
\begin{equation}
\frac{1}{M_*^2} \int \sqrt{-g}\, d^4x \, \partial_\mu R \, J^\mu,
\end{equation}
where $M_*$ denotes the cutoff scale of the effective theory, while $g$ and $R$ represent the determinant of the metric and the Ricci scalar, respectively.

Over the past decades, a wide variety of baryogenesis mechanisms have been proposed in order to explain the observed baryon asymmetry of the Universe. Early developments include Grand Unified Theory (GUT) baryogenesis scenarios \cite{Yoshimura1978,Weinberg1979}, followed by electroweak baryogenesis \cite{Kuzmin1985,Trodden1999}, which relies on sphaleron processes within the Standard Model and its extensions. Another important class of models is leptogenesis \cite{Fukugita1986,Buchmuller2005}, where an initial lepton asymmetry is generated and subsequently converted into a baryon asymmetry through sphaleron transitions. 

In addition, spontaneous baryogenesis mechanisms \cite{Cohen1987,DeSimone2017} introduce effective chemical potentials associated with time-dependent scalar fields, allowing baryon asymmetry generation even in near-equilibrium conditions. Gravitational baryogenesis, first proposed in \cite{Davoudiasl2004}, exploits a coupling between the derivative of the Ricci scalar and the baryon current, leading to a dynamical violation of CPT symmetry during cosmic expansion. This idea has been extensively generalized in modified theories of gravity, including $f(R)$, $f(T)$ and more recently extended frameworks \cite{Lambiase2013,Oikonomou2015,Oikonomou2017}. 

Furthermore, models involving scalar fields coupled to dark energy \cite{Li2002,Li2003} have provided new insights into the possible connection between baryogenesis and the late-time acceleration of the Universe. Recent studies have also explored baryogenesis in alternative gravitational settings and non-standard cosmological backgrounds \cite{Pizza2015}, highlighting the richness and diversity of possible mechanisms. Despite these advances, a complete and consistent explanation of the baryon asymmetry remains an open challenge, motivating the exploration of new theoretical frameworks.
	
\section{Field equations and cosmological dynamics}
We consider a modified gravity theory based on the non-metricity scalar $Q$ coupled to a scalar field $\phi$.  The total action is written as follows.

\begin{equation}
S = \int d^4x \sqrt{-g} \left[
\frac{1}{2\kappa} f(Q,\phi)
- \frac{1}{2}\partial_\mu \phi \partial^\mu \phi
- V(\phi)
+ \mathcal{L}_m\right].
\label{eq:action_baryogenesis_fQphi}
\end{equation}

Scalar-nonmetricity theories, in which the non-metricity scalar $Q$ is coupled to a scalar field, have been extensively investigated in the literature \cite{Jimenez2018, Jimenez2020, Bahamonde2021, Dialektopoulos2022}. These models provide a natural extension of symmetric teleparallel gravity and offer a rich framework for cosmological applications.
%


Varying the action (\ref{eq:action_baryogenesis_fQphi}) with respect to the metric tensor $g_{\mu\nu}$, we obtain the generalized field equations
\begin{equation}
-\frac{2}{\sqrt{-g}} \nabla_{\alpha}
\left( \sqrt{-g}\, f_Q P^{\alpha}_{\ \mu\nu} \right)
+ \frac{1}{2} g_{\mu\nu} f
+ f_{\phi}\, \partial_\mu \phi \partial_\nu \phi
= \kappa T_{\mu\nu},
\label{eq:field_equations_fQphi}
\end{equation}
where $f_Q = \partial f/\partial Q$, $f_{\phi} = \partial f/\partial \phi$, and $P^{\alpha}_{\ \mu\nu}$ is the superpotential tensor associated with non-metricity. Variation of the action with respect to the scalar field $\phi$ yields the modified Klein--Gordon equation ,in a spatially flat FLRW background, gives
\begin{equation}
\ddot{\phi} + 3H\dot{\phi} + V'(\phi)
- \frac{1}{2\kappa} f_{\phi}
= 0.
\label{eq:kg_equation_fQphi_baryogenesis}
\end{equation}

Assuming a flat FLRW metric, the modified Friedmann equation becomes
\begin{equation}
3H^2 f_Q =
\kappa \left( \rho + \frac{1}{2}\dot{\phi}^2 + V(\phi) \right)
- \frac{1}{2} f,
\label{eq:friedmann_fQphi_1}
\end{equation}

while the dynamical equation is given by
\begin{equation}
\dot{H} =
- \frac{\kappa}{2 f_Q}
\left( \rho + p + \dot{\phi}^2 \right).
\label{eq:raychaudhuri_fQphi}
\end{equation}

\section{f(Q,$\phi$) BARYOGENESIS}
In this framework, where we take into account a minimal coupling between matter, geometry, and the scalar field, we consider a CP-violating interaction term responsible for the baryon asymmetry of the Universe, of the form:
\begin{equation}
S_{\text{int}} = \frac{1}{M_*^2} \int \sqrt{-g}\, d^4x \; \partial_\mu f(Q , \phi)\, J^\mu.
\label{eq:interaction_baryogenesis_Qphi}
\end{equation}
Unlike previous studies, we consider a unified coupling between the non-metricity scalar and a dynamical scalar field inside the baryogenesis sector through $\partial_\mu f(Q,\phi) J^\mu$.

Since the gravitational sector depends on both $Q$ and $\phi$, i.e.
$f = f(Q,\phi)$, its spacetime derivative is given by the chain rule:
\begin{equation}
\partial_\mu f(Q,\phi)
= f_Q\, \partial_\mu Q + f_\phi\, \partial_\mu \phi,
\label{eq:chain_rule}
\end{equation}
where $f_Q=\partial f/\partial Q$ and $f_\phi=\partial f/\partial \phi$.

In a homogeneous and isotropic FLRW background, spatial derivatives vanish and Eq.~(\ref{eq:chain_rule}) reduces to
\begin{equation}
\dot{f}
= f_Q \dot{Q} + f_\phi \dot{\phi}.
\label{eq:fdot_general}
\end{equation}

For the non-metricity scalar, we have
\begin{equation}
Q = 6H^2,
\label{eq:Q_definition}
\end{equation}
which implies
\begin{equation}
\dot{Q} = 12H\dot{H}.
\label{eq:Qdot}
\end{equation}

Substituting Eq.~(\ref{eq:Qdot}) into Eq.~(\ref{eq:fdot_general}), we obtain
\begin{equation}
\dot{f}
= f_Q \cdot 12H\dot{H}
+ f_\phi \dot{\phi}.
\label{eq:fdot_final}
\end{equation}

The baryon asymmetry generated by the CP-violating interaction is given by
\begin{equation}
\frac{n_B}{s}
\simeq
-\frac{15 g_b}{4\pi^2 g_*}
\frac{1}{M_*^2 T_D}
\left(
f_Q \cdot 12H\dot{H}
+ f_\phi \dot{\phi}
\right).
\label{eq:ns_final}
\end{equation}

This expression represents the general form of baryogenesis in $f(Q,\phi)$ gravity, where both geometric ($Q$) and scalar field ($\phi$) contributions are present.

We assume in this work that thermal equilibrium holds throughout the relevant cosmological epoch. Under this assumption, the Universe is considered to evolve quasi-statically from one equilibrium state to another, with the energy density related to the temperature $T$ as
\begin{equation}
\rho = \frac{\pi^2}{30} g_* T^4,
\label{eq:energy_density_radiation}
\end{equation}
where $g_*$ denotes the effective number of relativistic (effectively massless) degrees of freedom.

In the framework of General Relativity, and by extension in the present modified gravity scenario, the matter content of the Universe is modeled as a perfect fluid characterized by a constant equation-of-state parameter defined as $w = p/\rho$, so that the pressure is given by $p = w\rho$.

In the following, we will consider specific forms of the function $f(Q,\phi)$ in order to investigate the baryogenesis mechanism. In particular, we will focus on the model

\subsection{ $ f(Q,\phi) = Q + \xi Q \phi^2$ }
We consider a nonminimally coupled scalar–nonmetricity model of the form $f(Q,\phi)=Q+\xi Q \phi^2$, which constitutes a natural extension of $f(Q)$ gravity by incorporating a scalar field nonminimally coupled 
to the nonmetricity scalar. The $f(Q)$ framework itself arises from the symmetric teleparallel formulation of gravity, where gravitation is attributed to nonmetricity rather than curvature or torsion 
\cite{Jimenez2018}. Such nonminimal couplings are well motivated from the perspective 
of scalar–tensor theories, which have been extensively studied in the context of modified gravity \cite{Harko2018}. In particular, analogous constructions involving couplings of the form 
$f(R,\phi)$ and $f(T,\phi)$ have been widely
investigated in the literature, providing viable frameworks to address cosmological issues such as inflation and late-time acceleration. Therefore, 
the model considered here represents a natural and theoretically well-grounded generalization of scalar–tensor theories within the nonmetricity-based gravitational paradigm \cite{Anagnostopoulos2021}. 
The equations \eqref{eq:kg_equation_fQphi_baryogenesis}, 
\eqref{eq:friedmann_fQphi_1}, and 
\eqref{eq:raychaudhuri_fQphi} 
take the following forms, respectively:
\begin{equation}
\ddot{\phi} + 3H\dot{\phi} + V'(\phi)
- \frac{\xi Q \phi}{\kappa} = 0,
\label{eq:kg_equation_specific_model}
\end{equation}
\begin{equation}
3H^2 (1 + \xi \phi^2) =
\kappa \left( \rho + \frac{1}{2}\dot{\phi}^2 + V(\phi) \right)
- \frac{1}{2} Q (1 + \xi \phi^2),
\label{eq:friedmann_specific_model}
\end{equation}
\begin{equation}
\dot{H} =
- \frac{\kappa}{2 (1 + \xi \phi^2)}
\left( \rho + p + \dot{\phi}^2 \right),
\label{eq:raychaudhuri_specific_model}
\end{equation}

In the radiation-dominated epoch relevant for baryogenesis, we assume that the scalar field evolves slowly compared to the Hubble expansion rate. This allows us to neglect the second time derivative of the scalar field, leading to a quasi-static approximation analogous to the slow-roll regime.

Assuming a power-law expansion of the form $a(t) = a_{0} t^\gamma$, the Hubble parameter reduces to $H = \gamma/t$, which implies that the nonmetricity scalar behaves as $Q = 6\gamma^2/t^2$. Under the slow-roll approximation, where the second derivative of the scalar field is negligible, and considering the regime in which the nonminimal coupling term dominates over the scalar potential contribution, the Klein-Gordon equation simplifies significantly and leads to a first-order differential equation for the scalar field. Solving this equation yields a power-law behavior for the scalar field given by
\begin{equation}
\phi(t) = \phi_0\, t^{\frac{2\xi \gamma}{\kappa}},
\label{eq:phi_solution_powerlaw}
\end{equation}
where $\phi_0$ is an integration constant. Consequently, the time derivative of the scalar field is obtained as
\begin{equation}
\dot{\phi}(t) =
\frac{2\xi \gamma}{\kappa}
\phi_0\, t^{\frac{2\xi \gamma}{\kappa} -1}.
\label{eq:phidot_solution_powerlaw}
\end{equation}
In the early Universe, where the Hubble parameter is large, the nonmetricity scalar $Q = 6H^2$ becomes significantly enhanced, which naturally amplifies the contribution of the nonminimal coupling term. As a consequence, the term $\frac{\xi Q \phi}{\kappa}$ can dominate over the scalar potential contribution. Moreover, for an exponential potential, the derivative $V'(\phi) \propto e^{-\lambda \phi}$ rapidly decreases as the scalar field evolves, further supporting the dominance of the coupling term. Such regimes, where geometric or coupling contributions dominate the scalar field dynamics, are commonly considered in modified gravity and cosmological models. Therefore, it is physically well-motivated to focus on the regime in which the nonminimal coupling drives the scalar field evolution \cite{Davoudiasl2004,Jimenez2018,Harko2018,Copeland1998,Nojiri2011}.

Substituting the expressions of the Hubble parameter $H = \gamma/t$, the equation of state $p = w\rho$, and the scalar field derivative $\dot{\phi}(t)$ into the dynamical equation \ref{eq:raychaudhuri_specific_model} , we obtain the following expression for the energy density:
\begin{equation}
\rho(t) =
\frac{1}{1+w}
\left[
\frac{2\gamma}{\kappa t^2}
\left(1 + \xi \phi_0^2 t^{\frac{4\xi \gamma}{\kappa}}\right)
-
\left(\frac{2\xi \gamma}{\kappa}\right)^2
\phi_0^2
t^{\frac{4\xi \gamma}{\kappa} -2}
\right].
\label{eq:rho_solution_powerlaw}
\end{equation}

The energy density receives contributions from both the standard cosmological term scaling as $t^{-2}$ and additional corrections induced by the nonminimal coupling and the scalar field dynamics.

By equating the modified energy density derived from the cosmological dynamics with the standard radiation energy density, one can explicitly determine the decoupling time $t_D$ as a function of the decoupling temperature $T_D$. In the leading-order approximation, where the dominant contribution scales as $t^{-2}$, this procedure yields the following expression:
\begin{equation}
t_D =
\left[
\frac{90\gamma^{2}}{\pi^2 g_* \kappa (1+w)}
\right]^{1/2}
\frac{1}{T_D^2}.
\label{eq:decoupling_time}
\end{equation}

The decoupling time is expressed as a function of the decoupling temperature by equating the modified energy density with the standard radiation energy density, leading to a relation of the form $t_D \propto T_D^{-2}$ in the high-temperature regime.Although the nonminimal coupling term dominates over the scalar potential in the Klein–Gordon equation, its contribution to the total energy density remains subdominant compared to the standard cosmological term. This justifies neglecting higher-order coupling corrections in the Friedmann sector while retaining their dynamical effects in the scalar field evolution.

Using Eq.(\ref{eq:decoupling_time}), we arrive at the final expression of the baryon-to-entropy ratio for the present  f($Q,\phi$) particular model

\begin{equation}
\frac{n_B}{s}
\simeq
\begin{aligned}
&\frac{15 g_b}{4\pi^2 g_*}
\frac{12\gamma^2}{M_*^2}
\left[
\frac{\pi^2 g_* \kappa (1+w)}{90\gamma^2}
\right]^{3/2}
T_D^5 \\
&\times\Bigg[
1
+ \xi \phi_0^2
\left(
\frac{90\gamma^2}{\pi^2 g_* \kappa (1+w)}
\right)^{\frac{2\xi \gamma}{\kappa}}
T_D^{-\frac{8\xi \gamma}{\kappa}}
- \frac{2\xi^2 \gamma}{\kappa}
\phi_0^2
\left(
\frac{90\gamma^2}{\pi^2 g_* \kappa (1+w)}
\right)^{\frac{2\xi \gamma}{\kappa}}
T_D^{-\frac{8\xi \gamma}{\kappa}}
\Bigg]
\end{aligned}
\label{eq:nb_over_s_final_full}
\end{equation}

In the radiation dominated phase $w = 1/3$, hence Eq.~\eqref{eq:nb_over_s_final_full} reduces to a simplified form where the equation of state parameter is fixed and the overall prefactor is modified accordingly. Therefore, the baryon-to-entropy ratio becomes
\begin{equation}
\left(\frac{n_B}{s}\right)_{\text{rad}}
\simeq
\frac{15 g_b}{4\pi^2 g_*}
\frac{12\gamma^2}{M_*^2}
\left[
\frac{4\pi^2 g_* \kappa}{270\gamma^2}
\right]^{3/2}
T_D^5
\Bigg[
1
+ \xi \phi_0^2
\left(
\frac{270\gamma^2}{4\pi^2 g_* \kappa}
\right)^{\frac{2\xi \gamma}{\kappa}}
T_D^{-\frac{8\xi \gamma}{\kappa}}
- \frac{2\xi^2 \gamma}{\kappa}
\phi_0^2
\left(
\frac{270\gamma^2}{4\pi^2 g_* \kappa}
\right)^{\frac{2\xi \gamma}{\kappa}}
T_D^{-\frac{8\xi \gamma}{\kappa}}
\Bigg].
\label{eq:nb_over_s_radiation}
\end{equation}

%

In order to obtain a baryon-to-entropy ratio compatible with observational constraints, $n_B/s$ , we fix the parameters as $M_* = 10^{12}\,\mathrm{GeV}$, $T_D = 2\times10^{16}\,\mathrm{GeV}$, $g_* = 106$, $g_b = 1$, and $\kappa = 1$. We further consider physically reasonable values of the expansion parameter $\gamma$, and we allow the coupling parameter $\xi$ to vary. As a result, the observed baryon asymmetry can be reproduced without requiring extreme fine-tuning of $\gamma$. The corresponding numerical values are summarized in Table below. In particular, for $\gamma = 0.3$, we obtain
$\frac{n_B}{s} \simeq 9.6 \times 10^{-11}$, which is in very good agreement with observations and practically equal to the  observed value $(n_B/s \simeq 9.42 \times 10^{-11})$. The corresponding numerical values are summarized below:





\begin{equation}
\begin{array}{|c|cccccc|}
\hline
\gamma & 0.60 & 0.50 & 0.40 & 0.30 & 0.20 & 0.10 \\
\hline
\xi & 18 & 15 & 12 & 11 & 10 & 8 \\
\hline
\frac{n_B}{s} & 
1.2\times10^{-11} & 
3.8\times10^{-11} & 
7.5\times10^{-11} & 
9.6\times10^{-11} & 
1.3\times10^{-10} & 
1.8\times10^{-10} \\
\hline
\end{array}
\end{equation}

\begin{figure}[h!]
	\centering
	\includegraphics[width=0.75\textwidth]{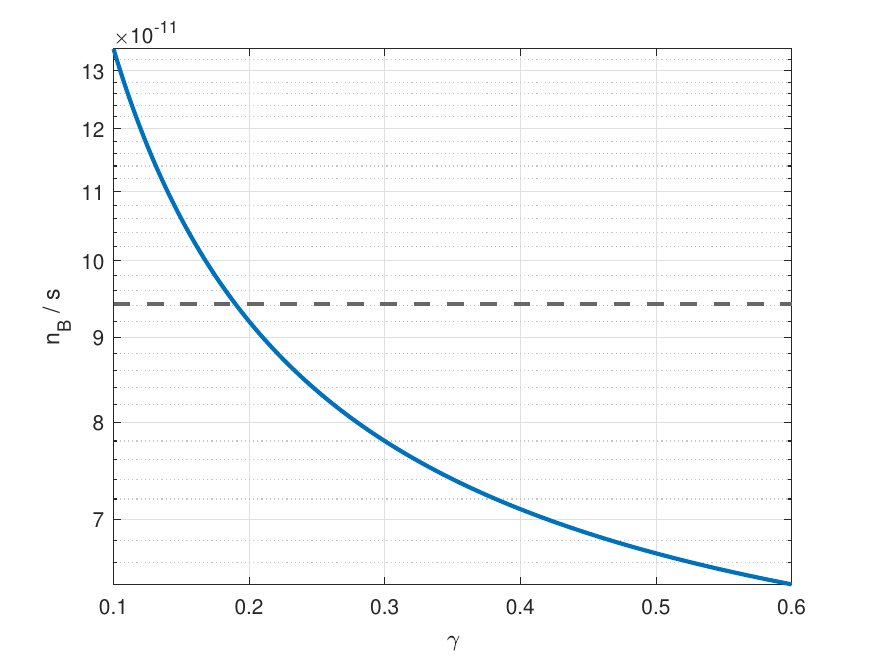}
	\caption{Evolution of the baryon-to-entropy ratio $n_B/s$ as a function of the coupling parameter $\xi$ within the model $f(Q,\phi)= Q + \xi Q \phi^2$. The parameters are fixed as $M_* = 10^{12}\,\mathrm{GeV}$, $T_D = 2\times10^{16}\,\mathrm{GeV}$, $g_* = 106$, $g_b = 1$, and $\kappa = 1$, while the expansion parameter is chosen as $\gamma = 0.3$. The horizontal dashed line represents the observed value $n_B/s \approx 9.42 \times 10^{-11}$. The model successfully reproduces the baryon asymmetry without requiring extreme fine-tuning of the parameters. In particular, for $\gamma = 0.3$, we obtain $n_B/s \simeq 9.6 \times 10^{-11}$, which is in excellent agreement with observational data.}
	\label{fig:nb_s_xi}
\end{figure}

The figure illustrates the evolution of the baryon-to-entropy ratio $ \frac{n_B}{s} $ as a function of the parameter $ \gamma $ within the framework of the model $ f(Q,\phi) = Q + \xi Q \phi^2 $. It is clearly observed that the curve exhibits a monotonically decreasing behavior with increasing $ \gamma $. For small values of $ \gamma $, the baryon-to-entropy ratio is relatively large, of the order of $ 10^{-10} $, while it gradually decreases toward values of the order of $ 10^{-11} $ as $ \gamma $ approaches higher values. This behavior reflects the strong dependence of baryogenesis on the cosmic expansion rate, since $ \gamma $ characterizes the evolution law of the scale factor.

Furthermore, the dashed horizontal line represents the observational constraint on the baryon-to-entropy ratio, $ \frac{n_B}{s} \sim 9 \times 10^{-11} $. The intersection between the theoretical curve and this observational value occurs around $ \gamma \approx 0.2 $, indicating that the model successfully reproduces the observed baryon asymmetry for this specific range of the parameter. This provides an important constraint on the allowed values of $ \gamma $ in the considered scenario.

From a physical point of view, the decrease of $\frac{n_B}{s}$ with increasing $\gamma$ can be interpreted as a consequence of a faster cosmic expansion, which reduces the efficiency of the baryogenesis mechanism. In this context, the nonminimal coupling term $ \xi Q \phi^2 $ plays a crucial role by modifying the dynamics of the nonmetricity scalar $ Q $ and its interaction with the scalar field $ \phi $, thereby directly affecting the generated baryon asymmetry.

Finally, the overall behavior of the curve, together with its compatibility with observational bounds, demonstrates that the model $ f(Q,\phi) = Q + \xi Q \phi^2 $ provides a viable framework for gravitational baryogenesis, with the parameter $ \gamma $ acting as a key quantity controlling the magnitude of the baryon-to-entropy ratio.

\subsection{$f(Q,\phi) = \alpha Q^n + \beta \phi Q$}

We consider a generalized scalar-nonmetricity model of the form $f(Q,\phi)=\alpha Q^n + \beta \phi Q$, which extends the standard $f(Q)$ gravity by introducing both a nonlinear function of the nonmetricity scalar and a direct coupling between the scalar field and $Q$. The term $\alpha Q^n$ represents a power-law modification of symmetric teleparallel gravity, which has been widely explored in the literature as a viable framework to describe the late-time cosmic acceleration without invoking a cosmological constant \cite{Lazkoz2019}. On the other hand, the linear coupling term $\beta \phi Q$ encodes an interaction between the scalar field and the gravitational sector, analogous to nonminimal couplings appearing in scalar–tensor theories and their extensions in $f(R)$ and $f(T)$ gravities \cite{Harko2018}. Such couplings naturally arise in effective field theory approaches and can significantly modify the cosmological dynamics, leading to rich phenomenology at both early- and late-time evolution. Therefore, the present model provides a unified framework that combines nonlinear geometric corrections with scalar–gravity interactions within the nonmetricity-based description of gravitation \cite{Anagnostopoulos2021}. The Klein--Gordon equation
(\ref{eq:kg_equation_fQphi_baryogenesis}), the modified Friedmann equation (\ref{eq:friedmann_fQphi_1}), and the Raychaudhuri equation (\ref{eq:raychaudhuri_fQphi}) respectively reduce to
\begin{equation}
\ddot{\phi} + 3H\dot{\phi} + V'(\phi)
- \frac{\beta Q}{2\kappa}
= 0,
\label{eq:kg_equation_powerlaw_model}
\end{equation}
\begin{equation}
3H^2 \left( \alpha n Q^{n-1} + \beta \phi \right) =
\kappa \left( \rho + \frac{1}{2}\dot{\phi}^2 + V(\phi) \right)
- \frac{1}{2} \left( \alpha Q^n + \beta \phi Q \right),
\label{eq:friedmann_powerlaw_model}
\end{equation}
\begin{equation}
\dot{H} =
- \frac{\kappa}{2 \left( \alpha n Q^{n-1} + \beta \phi \right)}
\left( \rho + p + \dot{\phi}^2 \right).
\label{eq:raychaudhuri_powerlaw_model}
\end{equation}
	
We consider a power-law expansion of the form $a(t)=a_0 t^\gamma$, which implies $H=\gamma/t$ and $Q=6H^2=6\gamma^2/t^2$. Under the slow-roll approximation where the coupling term dominates over the scalar potential, the Klein--Gordon equation reduces to a first-order differential equation whose solution yields the scalar field and its time derivative as
\begin{equation}
\phi(t) = \frac{\beta \gamma}{\kappa} \ln t + \phi_0,
\label{eq:phi_solution_log_model}
\end{equation}
\begin{equation}
\dot{\phi}(t) = \frac{\beta \gamma}{\kappa} \frac{1}{t},
\label{eq:phidot_solution_log_model}
\end{equation}

We determine the energy density by substituting the power-law expansion, the expression of the nonmetricity scalar, and the scalar field solutions into the Raychaudhuri equation. This leads to
\begin{equation}
\rho(t) =
\frac{1}{1+w}
\left[
\frac{2\gamma}{\kappa t^2}
\left(
\alpha n \left(\frac{6\gamma^2}{t^2}\right)^{n-1}
+ \beta \left( \frac{\beta \gamma}{\kappa} \ln t + \phi_0 \right)
\right)
-
\left(\frac{\beta \gamma}{\kappa}\right)^2 \frac{1}{t^2}
\right].
\label{eq:rho_powerlaw_log_model}
\end{equation}

Under the high-temperature approximation where the nonminimal geometric contribution dominates over the scalar field term, the energy density reduces to a power-law form.The decoupling time is obtained by equating the modified energy density (\ref{eq:rho_powerlaw_log_model}) with the standard radiation energy density. This leads to
\begin{equation}
t_D =
\left[
\frac{60\, \alpha n\, \gamma\, (6\gamma^2)^{\,n-1}}
{\pi^2 g_* \kappa (1+w)}
\right]^{\frac{1}{2n}}
T_D^{-\frac{2}{n}}.
\label{eq:tD_powerlaw_general}
\end{equation}

The obtained expression for the decoupling time exhibits a clear dependence on the power-law index $n$ of the gravitational function. In the particular case $n=1$, one recovers the standard scaling $t_D \propto T_D^{-2}$, which corresponds to the usual behavior found in General Relativity. However, for $n>1$, the decoupling time evolves more slowly with temperature, indicating a delayed thermal history compared to the standard scenario. In contrast, for $n<1$, the evolution becomes faster, leading to an earlier decoupling epoch. This deviation from the standard scaling highlights the direct impact of the modified gravity sector on the thermal evolution of the Universe and provides a distinctive signature of the $f(Q,\phi)$ framework in the context of baryogenesis. Although the scalar field evolves logarithmically with time, its contribution remains subdominant compared to the geometric term proportional to $Q^{n-1}$. In the early Universe ($t \to 0$), the nonmetricity scalar scales as $Q \sim t^{-2}$, leading to a much faster growth of the geometric contribution. 

By substituting the expressions of the scalar field, the nonmetricity scalar, and the Hubble parameter, together with the decoupling time $t_D$, into Eq.~(\ref{eq:ns_final}), we obtain the following explicit expression for the baryon-to-entropy ratio in terms of the temperature $T_D$ and the model parameters:

\begin{equation}
\begin{aligned}
\frac{n_B}{s}
\simeq\;
& \frac{15 g_b}{4\pi^2 g_*}
\frac{1}{M_*^2 T_D}
\left[
\frac{60\, \alpha n\, \gamma\, (6\gamma^2)^{n-1}}
{\pi^2 g_* \kappa (1+w)}
\right]^{-\frac{3}{2n}}
T_D^{\frac{6}{n}}
\\[0.3cm]
& \times
\Bigg\{
12\gamma^2
\Big[
\alpha n (6\gamma^2)^{n-1}
\left[
\frac{60\, \alpha n\, \gamma\, (6\gamma^2)^{n-1}}
{\pi^2 g_* \kappa (1+w)}
\right]^{-\frac{n-1}{n}}
T_D^{\frac{4(n-1)}{n}}
\\
& \qquad\qquad
+ \beta
\left(
\frac{\beta \gamma}{\kappa}
\ln\!\left[
\left(
\frac{60\, \alpha n\, \gamma\, (6\gamma^2)^{n-1}}
{\pi^2 g_* \kappa (1+w)}
\right)^{\frac{1}{2n}}
T_D^{-\frac{2}{n}}
\right]
+ \phi_0
\right)
\Big]
\\
& \qquad
- \frac{6\beta^2 \gamma^3}{\kappa}
\Bigg\}.
\end{aligned}
\label{eq:nb_s_clean}
\end{equation}

In the radiation-dominated epoch, where $w = \frac{1}{3}$, the baryon-to-entropy ratio given in Eq.~(\ref{eq:nb_s_clean}) reduces to the following form:

\begin{equation}
\begin{aligned}
\frac{n_B}{s}
\simeq\;
& \frac{45 g_b}{4\pi^2 g_*}
\frac{1}{M_*^2 T_D}
\left[
\frac{45\, \alpha n\, \gamma\, (6\gamma^2)^{n-1}}
{\pi^2 g_* \kappa}
\right]^{-\frac{3}{2n}}
T_D^{\frac{6}{n}}
\\[0.3cm]
& \times
\Bigg\{
12\gamma^2
\Big[
\alpha n (6\gamma^2)^{n-1}
\left[
\frac{45\, \alpha n\, \gamma\, (6\gamma^2)^{n-1}}
{\pi^2 g_* \kappa}
\right]^{-\frac{n-1}{n}}
T_D^{\frac{4(n-1)}{n}}
\\
& \qquad
+ \beta
\left(
\frac{\beta \gamma}{\kappa}
\ln\!\left[
\left(
\frac{45\, \alpha n\, \gamma\, (6\gamma^2)^{n-1}}
{\pi^2 g_* \kappa}
\right)^{\frac{1}{2n}}
T_D^{-\frac{2}{n}}
\right]
+ \phi_0
\right)
\Big]
\\
& \qquad
- \frac{6\beta^2 \gamma^3}{\kappa}
\Bigg\}.
\end{aligned}
\label{eq:nb_s_radiation_simplified}
\end{equation}

\begin{figure}[h!]
	\centering
	\includegraphics[width=0.75\textwidth]{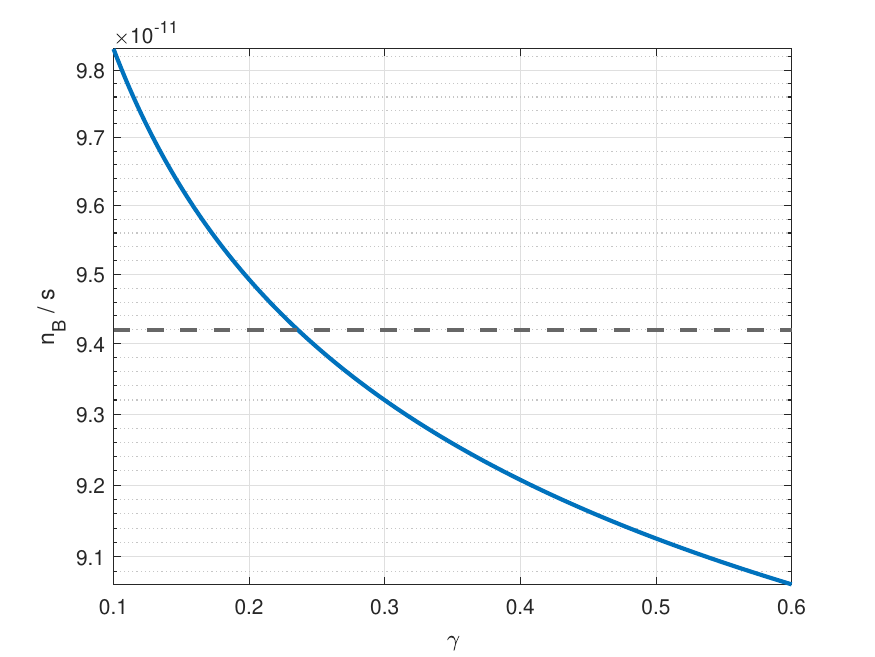}
	\caption{Evolution of the baryon-to-entropy ratio $n_B/s$ as a function of the parameter $\gamma$ within the $f(Q,\phi)=\alpha Q^n + \beta \phi Q$ model. The solid curve represents the theoretical prediction, while the horizontal dashed line indicates the observed value $n_B/s \approx 9.2 \times 10^{-11}$. For the parameter values $\alpha = 8 \times 10^{-3}$ and $\beta = 7 \times 10^{-2}$, the model yields $n_B/s = 9.1 \times 10^{-11}$, showing a very good agreement with observational constraints.}
	\label{fig:nb_s_gamma}
\end{figure}

\noindent
For the fixed parameter values 
$M_* = 10^{12}\,\mathrm{GeV}$, 
$T_D = 2\times10^{16}\,\mathrm{GeV}$, 
$g_* = 106$, 
$g_b = 1$, 
$\kappa = 1$, 
and by choosing $n = 2$, $\gamma = 1$, and $\phi_0 = 1$, 
we numerically evaluate the baryon-to-entropy ratio $n_B/s$ 
for different values of the parameters $\alpha$ and $\beta$. 
We find that the observed range $10^{-11} \lesssim n_B/s \lesssim 10^{-10}$ 
can be achieved for specific combinations of these parameters, as summarized below. For the parameter values $\alpha = 8 \times 10^{-3}$ and $\beta = 7 \times 10^{-2}$, the model predicts a baryon-to-entropy ratio $n_B/s = 9.1 \times 10^{-11}$, which is in very good agreement with the observational value. This result indicates that the considered model is capable of successfully reproducing the baryon asymmetry of the Universe.

\begin{center}
	\begin{tabular}{c|c|c|c|c|c|c}
		$\alpha$ 
		& $1.0\times10^{-3}$ 
		& $2.0\times10^{-3}$ 
		& $5.0\times10^{-3}$ 
		& $1.5\times10^{-2}$ 
		& $8.0\times10^{-3}$ 
		& $1.0\times10^{-2}$ \\
		
		\hline
		
		$\beta$ 
		& $2.0\times10^{-2}$ 
		& $3.0\times10^{-2}$ 
		& $5.0\times10^{-2}$ 
		& $1.0\times10^{-1}$ 
		& $7.0\times10^{-2}$ 
		& $8.0\times10^{-2}$ \\
		
		\hline
		
		$\frac{n_B}{s}$ 
		& $1.3\times10^{-11}$ 
		& $2.7\times10^{-11}$ 
		& $6.5\times10^{-11}$ 
		& $8.7\times10^{-11}$ 
		& $9.1\times10^{-11}$ 
		& $1.2\times10^{-10}$ \\
	\end{tabular}
\end{center}

\noindent

The figure illustrates the evolution of the baryon-to-entropy ratio $\frac{n_B}{s}$ as a function of the parameter $\gamma$ within the framework of the $f(Q,\phi) = \alpha Q^n + \beta \phi Q$ gravity model. It is observed that $\frac{n_B}{s}$ decreases monotonically as $\gamma$ increases, while remaining of the order of $10^{-11}$, which is consistent with cosmological observations. The horizontal dashed line represents the observed value $\left(\frac{n_B}{s}\right)_{\text{obs}} \approx 9.2 \times 10^{-11}$, and the model predictions intersect this value around $\gamma \approx 0.2$--$0.25$, indicating good agreement with observational constraints. Physically, the parameter $\gamma$ is related to the expansion dynamics of the Universe, typically through a power-law behavior of the scale factor. For lower values of $\gamma$, the expansion rate is slower, allowing for enhanced baryon production, whereas for higher values of $\gamma$, the faster expansion leads to a stronger dilution of baryon asymmetry, resulting in a decrease of $\frac{n_B}{s}$. Furthermore, the structure of the modified gravity model plays a crucial role in this behavior. The nonlinear term $Q^n$ with $n < 0$ enhances the gravitational effects at low values of $Q$, while the coupling term $\beta \phi Q$ introduces scalar field dynamics that influence the baryogenesis process. Overall, the model successfully reproduces the correct order of magnitude for the baryon asymmetry and exhibits a clear dependence on the parameter $\gamma$, which can be constrained through observational data.

The results are ordered in increasing values of $n_B/s$, showing that the baryon asymmetry grows with increasing $\alpha$ and $\beta$, and remains consistent with observational bounds within the range $\alpha \sim 10^{-3} - 10^{-2}$ and $\beta \sim 10^{-2} - 10^{-1}$.

\section{Conclusion}

In this work, we have investigated gravitational baryogenesis in the framework of scalar–nonmetricity theories, focusing on two specific models, namely $f(Q,\phi)=Q+\xi Q \phi^2$ and $f(Q,\phi)=\alpha Q^n + \beta \phi Q$. These models represent natural extensions of $f(Q)$ gravity by incorporating nonminimal couplings between the scalar field and the nonmetricity scalar, leading to rich cosmological implications.

For the first model, $f(Q,\phi)=Q+\xi Q \phi^2$, we have analyzed the evolution of the baryon-to-entropy ratio $\frac{n_B}{s}$ as a function of the expansion parameter $\gamma$. The results show a clear monotonically decreasing behavior of $\frac{n_B}{s}$ with increasing $\gamma$, highlighting the crucial role of the cosmic expansion rate in the baryogenesis mechanism. In particular, for small values of $\gamma$, the baryon asymmetry is enhanced, reaching values of order $10^{-10}$, while it decreases toward $10^{-11}$ for larger $\gamma$. The comparison with observational constraints, $\frac{n_B}{s} \sim 9 \times 10^{-11}$, reveals that the model successfully reproduces the observed baryon asymmetry around $\gamma \approx 0.2$--$0.3$. 

By fixing the parameters to physically motivated values, $M_* = 10^{12}\,\mathrm{GeV}$, $T_D = 2\times10^{16}\,\mathrm{GeV}$, $g_* = 106$, $g_b = 1$, and $\kappa = 1$, we have shown that the observed baryon asymmetry can be obtained without requiring severe fine-tuning. For instance, for $\gamma = 0.3$, we obtain $\frac{n_B}{s} \simeq 9.6 \times 10^{-11}$, which is in excellent agreement with the observed value. These results demonstrate that the nonminimal coupling term $\xi Q \phi^2$ plays a fundamental role in controlling the efficiency of baryogenesis through its impact on the gravitational dynamics.

For the second model, $f(Q,\phi)=\alpha Q^n + \beta \phi Q$, we have explored the combined effects of nonlinear geometric corrections and scalar–gravity interactions. By fixing $n=2$, $\gamma=1$, and $\phi_0=1$, and varying the parameters $\alpha$ and $\beta$, we have shown that the baryon-to-entropy ratio can naturally lie within the observational range $10^{-11} \lesssim n_B/s \lesssim 10^{-10}$. In particular, for $\alpha = 8 \times 10^{-3}$ and $\beta = 7 \times 10^{-2}$, we obtain $\frac{n_B}{s} = 9.1 \times 10^{-11}$, in very good agreement with observations. The results indicate that the baryon asymmetry increases with increasing values of $\alpha$ and $\beta$, remaining consistent with observational bounds for $\alpha \sim 10^{-3} - 10^{-2}$ and $\beta \sim 10^{-2} - 10^{-1}$.

Furthermore, the analysis of the dependence of $\frac{n_B}{s}$ on $\gamma$ in this model confirms a similar decreasing behavior, with the theoretical predictions intersecting the observed value around $\gamma \approx 0.2$--$0.25$. This reinforces the interpretation that faster cosmic expansion leads to a suppression of baryon production due to stronger dilution effects. The presence of the nonlinear term $Q^n$ and the coupling $\beta \phi Q$ significantly enriches the dynamics, allowing for a more flexible and realistic description of baryogenesis.

In conclusion, both models considered in this work provide viable and consistent frameworks for gravitational baryogenesis within the context of nonmetricity-based gravity. The results highlight the importance of the expansion parameter $\gamma$ and the coupling constants in determining the magnitude of the baryon asymmetry. These findings open new perspectives for exploring the interplay between modified gravity and particle physics in the early Universe, and they suggest that scalar–nonmetricity theories can serve as promising candidates for explaining the origin of the baryon asymmetry of the Universe.

\end{document}